\documentclass[%
 aip,
 amsmath,amssymb,
 reprint,%
]{revtex4-1}
\usepackage[utf8]{inputenc}
\usepackage{lmodern}		
\usepackage{cmap}			

\usepackage{lipsum}
\usepackage{calc}
\usepackage{xifthen}
\usepackage[]{algorithm2e}
\usepackage[table,xcdraw]{xcolor}
\usepackage{xr}
\usepackage{amssymb}
\usepackage{amsmath}
\usepackage{multirow}
\usepackage{todonotes}
\usepackage{graphics,graphicx}
\usepackage{xspace}
\usepackage{dsfont}
\usepackage{bm}
\usepackage{mathtools}
\usepackage{hyperref}
\usepackage{verbatim}

\usepackage{color, soul}
\newcommand{\figref}[1]{Fig.\ref{#1}}
\newcommand{\cv}{$c_\lambda$}
\newcommand{\eqreff}[1]{Eq.\eqref{#1}}

\hypersetup{
	colorlinks,%
	citecolor=blue,%
	filecolor=blue,%
	linkcolor=blue,%
	urlcolor=blue
}

\begin{document}
\title{Characterization and comparison of large directed graphs through the spectra of the magnetic Laplacian}
\author{Bruno Messias F. de Resende}
\email{messias.physics@gmail.com}
\affiliation{
	Physics Institute of São Carlos, University of São Paulo,  São Carlos, SP 13566-590, Brazil}

\author{Luciano da F. Costa}
\affiliation{
	Physics Institute of São Carlos, University of São Paulo,  São Carlos, SP 13566-590, Brazil}
\date{\today}
\begin{abstract}


	In this paper we investigated the possibility to use the magnetic Laplacian to characterize directed graphs (a.k.a. networks). Many interesting results are obtained, including the finding that community structure is related to rotational symmetry in the spectral measurements for a type of stochastic block model. Due the hermiticity property of the magnetic Laplacian we show here how to scale our approach to larger networks containing hundreds of thousands of nodes using the Kernel Polynomial Method (KPM). We also propose to combine the KPM with the Wasserstein metric in order to measure distances between networks even when these networks are directed, large and have different sizes, a hard problem which cannot be tackled by previous methods presented in the literature. In addition, our python package  is publicly available at \href{https://github.com/stdogpkg/emate}{github.com/stdogpkg/emate}. The codes can run in both CPU and GPU and can estimate the spectral density and related trace functions, such as entropy and  Estrada index, even in directed or undirected networks with million of nodes.
\end{abstract}

\maketitle
\begin{quotation}
	The Laplacian operator of a directed network is not Hermitian. This property hampers the interpretation of the spectral measurements and restricts the use of computational methods developed in network science. In this work, we propose a framework and novel measures based on the spectrum of the magnetic Laplacian to study  directed networks. By using the properties of circulant matrices, we show analytically that novel measurements are able to grasp information about the structure of directed networks. It shows that the number of modular structures in  networks is related to the rotational symmetry of the spectrum, and therefore can contribute to characterize the parameters of the directed networks.  To infer the generative parameters of networks, we propose the application of the Wasserstein metric to measure the distance between the spectra of the magnetic Laplacian, allowing networks to be compared.   All the proposed methods depend on the diagonalization of the magnetic Laplacian operator, which implies a high computational cost. Therefore, the calculations can become unfeasible. To overcome this limitation, we implemented the Kernel Polynomial Method (KPM) using TensorFlow package. This method approximates the spectrum density of Hermitian matrices with a lower computational cost, allowing the spectral characterization of large directed networks containing hundreds of thousands of nodes.

\end{quotation}

\section{Introduction}

In the seminal work \textit{Can one hear the shape of a drum?}~\cite{Kac1966}  Mark Kack discusses the relationship between a membrane and the set of eigenvalues (spectrum) of the Laplacian operator. However, this relationship was identified to be not unique~\cite{RevModPhysShapeDrum2010}, in the sense that two distinct membranes (non-isometric manifolds) can have the same spectrum. Nevertheless, despite such degeneracies, spectral information can provide valuable insights about the real world.  For instance, spectral geometry has been used to study physical phenomena such as quantum gravity~\cite{Aasen2013} and provided the basis for developing algorithms in computer science~\cite{isospectralization2018}. 

Although the analysis of continuous regions such as those considered by Kack remains an interesting issue, several phenomena in nature and society need to be modeled in terms of discrete structures such as networks.  In this case, we can adapt Kack's question as \textit{Can one hear the shape of a network?}  The answer to this question is analogous to what has been verified for the original question, i.e.,~two nonisomorphic networks can share the same spectrum~\cite{cvetkovic1971graphs, vanDam2003}.  Despite such a limitation, the spectral approach to discrete structures can still be useful in some practical and theoretical problems\cite{Sarkar2018, Wang2017}.  An example of a spectral approach that has been applied to characterize networks is the von-Neumann entropy~\cite{Ginestra2009, Dehmer2011,pre2015}. 

More recently, the concept of entropy of a graph has been used to measure the similarity between two given networks~\cite{prx2016}.  Examples of this approach  include the entropic similarity applied to the inference of parameters of network models~\cite{prx2016, preModelFitEntropy2018}. However, this measure cannot be immediately extended to directed networks and, as has been shown in~\cite{PREPereira2015}, the directed edges have substantial implications in dynamics on graphs. 
 
In addition,  the entropic similarity depends on the product of the matrices associated with the given networks.  This implies that this similarity measurement is not invariant with respect to permutations of the indices associated with the nodes.  Given these dependencies, such measurements are not well defined when the nodes cannot be associated with fixed indices.

	Given that directed networks can accurately model several real-world problems, it is essential to develop new methodologies capable of dealing with network directionality.  	
An immediate difficulty implied by this scenario is that the associated Laplacian operator will often have complex values, because the adjacency matrix associated with the directed networks is non-symmetric.  A promising approach to address this problem consists of studying directed complex networks while considering their magnetic Laplacian operator~\cite{Berkolaiko2013,preMagCommunity2017, graphSignal2020}. As an example, in~\cite{preMagCommunity2017} the authors showed that the results of community detection algorithms could be improved by considering the magnetic Laplacian associated with the directed network.	

In this work, we show that the magnetic Laplacian approach can be used to characterize complex networks, including those	
with hundreds of thousands of nodes. By characterization, we mean that measurements taken from this operator contribute to identify the network model responsible for generating a given network, as well as performing the inference of parameters responsible for generating a given specific network configuration. 	
Several results were obtained.  First, for simpler models (i.e., modular regular networks),  the number of modular structures is related to the specific heat rotational symmetry.   Subsequently, we showed that these spectral measurements combined with the Wasserstein distance between spectral densities~\cite{Kantorovitch42, Bogachev2012, Peyr2019}, can provide valuable contributions to infer the original parameters used for getting those networks, with relative errors smaller than $1\%$.

\section{Methods}
 \begin{figure*}[!htb]
   \includegraphics[width=\textwidth]{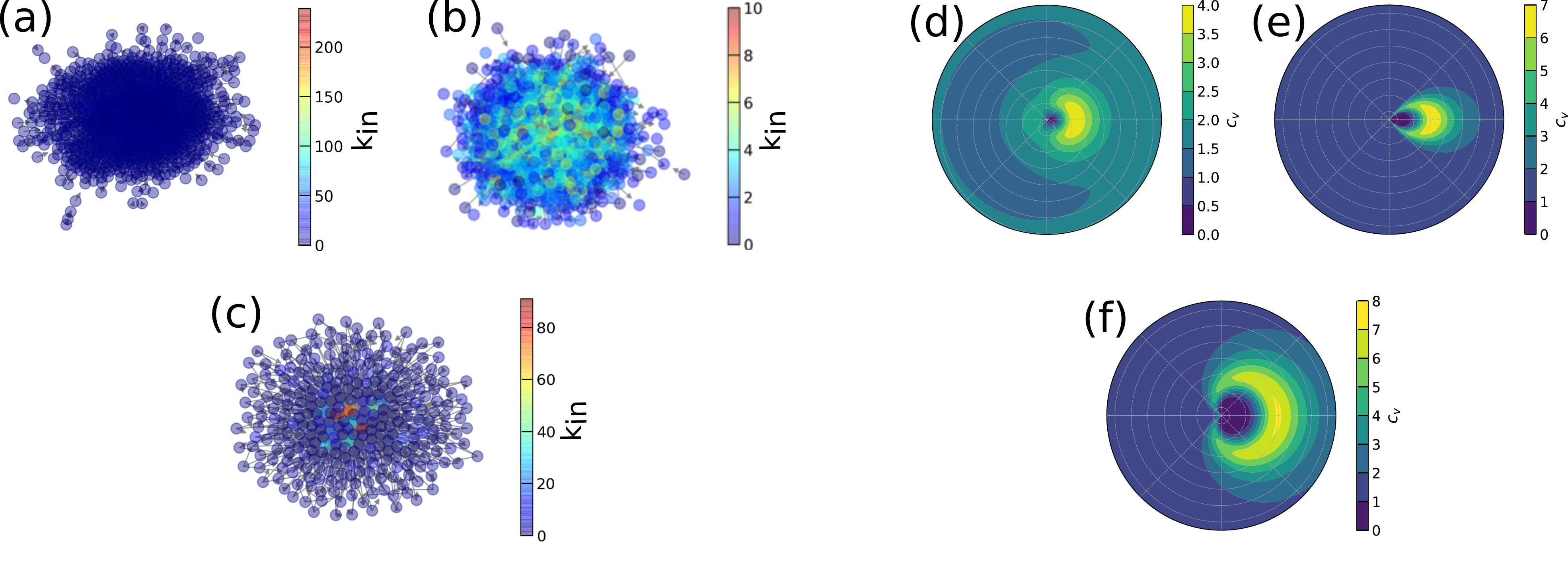}
 	  \caption{ In (a), (b) and (c) we have a SF, ER and BA network. The color maps, $k_{in}$ is the indegree of a given node. In (d), (e) and (f) it is shown the specific heat in terms of the charge $2\pi q$ (polar coordinates) and temperature (radial coordinate) for a Bollobas et al.~scale-free network \cite{bollobas2003directed},  ER , and BA network respectively. The parameters used to generate those networks were $|V|=1000$; the edge probability for ER was $p=0.003$; the  number of outgoing edges for BA network was $m=3$. The temperature range and charge are uniformly sampled form interval $[0.01, 0.15]$  and $[0, 1/2]$ with $30$ points each. As can be noted the $c_\lambda$ shows a specific pattern for each network. This \emph{fingerprint pattern} for each network explains why the SOM (Self-Organization Map) was so successful in the task of organizing networks belonging to the same classes onto the same groups using only the specific heat, without any knowledge about that classes.  It follows from  \eqreff{eqGamma} that the eigenvalues, and therefore \cv \ \,  are symmetric with respect to the addition of integer values to the charge $\gamma_q = \gamma_{q+j}\ \  \forall \ \ j  \in \mathbb{Z}$, reflecting in the bilateral symmetry with respect to the horizontal axis in (d), (e) and (f).}
  	\label{figNets2Cv}
   \end{figure*}
 
\subsection{Magnetic Laplacian}

A directed network can be expressed by a tuple $G=(V, E, w)$, where $V$ is the set of vertices, and $|V|$ stands for the number of the vertices; $E$ is the set of edges such that for each $u, v \in V$  the ordered tuple $e = (u, v) \in E$ assigns a directed edge from vertex $u$ to $v$ and  $w: E\to \mathbb{R}$. A directed network can be associated with an undirected counterpart  $G^{(s)}=(V , E^{(s)}, w^{(s)})$,  where $w^{(s)}(u, v) =\frac{w(u, v) + w(v, u)}{2}$.   However,  the directionality of  $G$ is lost in $G^{(s)}$. 

In order to preserve the Hermiticity and the information about directionality , define   $\gamma$, as 
$
\gamma: E \to  \mathcal{G},
$
where $\mathcal{G}$  is a group, such that  $\gamma(u, v )^{-1} = \gamma(v, u)$, choosing  $\mathcal{G} = U(1)$  and expressing $\gamma$  as
\begin{align}
\gamma_q(u, v) = \exp( 2  \pi i  q f(u,v)),
\label{eqGamma}
\end{align}
where $q \in [0, 1]$ and   $f(u, v) = w(u, v) - w(v, u)$  represents the  flow in a given vertex $u$ due to another vertex $v$.  

The symmetric network equipped with $\gamma_q$ has information about directed edges and, at the same time, the adjacency matrix is Hermitian.

Now, we consider the following operator, associated with $(G^{(s)}, \gamma_q)$ where $\odot$ is the Hadamard product:
\begin{align}
\mathbf L_q =\mathbf  D -\boldsymbol  \Gamma_q \odot\mathbf  W^{(s)},
\end{align}
where $\mathbf  D $  is the  degre matrix which contains the node degrees along its main diagonal; $[\boldsymbol  \Gamma_q]_{u,v}=[\boldsymbol  \Gamma_q^{\dagger}]_{v,u}=\gamma_q(u, v) $   and $[W^{(s)}]_{u,v} =[W^{(s)}]_{v,u} = w^{(s)}(u, v)$  .

It is interesting to observe that this operator corresponds to the magnetic Laplacian~\cite{colin2013magnetic,Berkolaiko2013}, $L_q$. The reason for the term magnetic is that the operator can be used to describe the phenomenology of a quantum particle subject to the action of a magnetic field~\cite{lieb1993fluxes}. Due to this physical context, the  parameter $q$ is named charge.

By construction, $\mathbf  D $ and $\mathbf  W^{(s)}$ are both symmetric and  $\boldsymbol  \Gamma_q$  is Hermitian. Consequently,  $\mathbf L_q $ is  Hermitian.  In addition, it is sometimes convenient to use a normalized version of $\mathbf L_q $, which is  given by

\begin{align}
\mathbf  H_q  
= 
\sqrt{\mathbf D^{-1}}\mathbf L_q \sqrt{\mathbf D^{-1}},
\label{eqNormedH}
\end{align}
where the $\mathbf  H_q$ is defined only if   the network is at least  weakly connected.

A given eigenvector of $\mathbf H_q$, $|\psi_{l,q}\rangle\ \in \mathbb{C}^{|V|} $,  can be obtained as solution of
\begin{align}
\mathbf H_q |\psi_{l,q}\rangle = \lambda_{l,q}|\psi_{l,q}\rangle
\label{eqEigenVec}
\end{align}

where $\lambda_{l,q}\in \mathbb R$ and $\lambda_{1,q}\le\lambda_{l,q}\leq\dots\leq\lambda_{|V|,q}$

It is possible to enhance the analogy with physical systems by including a temperature parameter $T \in \mathbb{R}_{+}$.  By using this parameter, the network properties can be studied from the statistical mechanics viewpoint.

Here, we  adopted the Boltzmann-Gibbs statistical mechanics formulation as a means to associate the  partition function

\begin{align}
Z(T, q)= \sum\limits_{l=1}^{|V|}e^{-\frac{\lambda_{l,q}}{T}}
\label{eqPartition}
\end{align}
with $G$.

By using~\eqreff{eqPartition}, the expected value at temperature T  of a  operator  $O$ can be expressed in terms of its eigenvalues $\{o_l\}$  as

\begin{align}
\langle O \rangle = \frac{1}{Z(T, q)}\sum\limits_{l=1}^{|V|} e^{-\frac{\lambda_{l,q}}{T}}o_l.
\label{eqExpectedValue}
\end{align}
	In this work, we  use~\eqreff{eqExpectedValue} to define the measure of specific heat,  $c_\lambda$,  associated with a network.  This novel measurement is given by
\begin{align}
c_\lambda(q, T) = \frac{\langle  H_q^2 \rangle- \langle  H_q \rangle^2}{T^2}.
\label{eqDefSpecifcHeat}
\end{align}

The~\eqreff{eqDefSpecifcHeat} has two free parameters, namely $q$ and  $T$. Because of this free choice of parameters and, owing to the fact that we have a rotation associated with directed edges ($\gamma_q$), we  plot  $c_\lambda$  in two dimensions , setting $2 \pi q$ as the polar coordinate, and $T$ as the radial one. 	
Regarding the interpretation and justification of physics-related quantities such as the specific heat it is directly related to the variance of the eigenvalue spectrum.  As a consequence, that  quantity  provides a signature of the spectrum properties, contributing to the characterization of the network structure.

\subsection{ Directed  modular networks}
\label{refSupMatNflux}
	In this work, we resort to a type of directed stochastic block model~\cite{} in order to obtain a good control of the network properties such as community size, and also because of its potential for facilitating analytical studies.  The adopted stochastic block model networks were obtained as follows	
\begin{enumerate}
\item Split the set $V$ onto $N_f$ equal-size sets ($f_1, f_2,\dots,f_{N_f}$).
\item For each $u, v \in f_i$ create a directed edge $(u, v)$ with probability $p_c$.
\item For each $u \in f_i$ and a $v \in f_{i+1}$ (assuming $f_{N_f+1}=f_1$), create a directed edge $(u, v)$ with probability $p_d$. 
\end{enumerate}

\subsection{Spectral entropy of directed networks}

Recent works reported how to use entropic measurements to quantify the similarity between two undirected networks~\citep{prx2016,preModelFitEntropy2018}.  The entropy of a network is derived from the usual Laplacian spectrum (all eigenvalues are real).  By contrast, these measurements cannot be used in the case of directed networks because the adjacency matrix is not Hermitian. However,  the magnetic Laplacian methodology yields a Hermitian operator
$H_q$,  which is here used to define an entropic measurement for directed networks.

Recall that a quantum system at finite temperature, $T$, is defined by its respective density matrix, $\rho(T)$~\cite{blum2012density}. For a network $G$ and charge $q$, this operator can be expressed in terms of the eigenvalues and eigenvectors associated to  $H_q$ as

\begin{align}
\boldsymbol \rho_q(T)
\frac{1}{Z(T,q)}\sum\limits_{l=1}^{|V|} e^{-\frac{\lambda_{q, l}}{T}}
|\psi_{l,q}\rangle \langle\psi_{l,q}|.
\label{eqDensity}
\end{align}

The previously defined density matrix can be used in order to define measurements associated with a directed (or undirected) network. 
For instance, by using the previous definition, the concepts of spectral entropy of a network can be extended for the directed case by using the following equation

\begin{align}
S(G, q, T) = \mathrm{Tr}\left[
	  \boldsymbol{ \rho}_q(T)\mathrm{Log}\boldsymbol{\rho}_q(T)\right],
      \label{eqEnt}
\end{align}
where $\mathrm{Log}$ is the matrix logarithm and $\mathrm{Tr}$ corresponds to the trace operation. 

Given the definition of spectral entropy, we can extend the entropic dissimilarity between two directed networks,   $\tilde G$  and $G$, as 

\begin{align}
S_d(\tilde G, G, q, T) = 
S(\tilde G, q, T)-\mathrm{Tr}\left[\boldsymbol{\tilde { \rho}}_q(T) \mathrm{Log} \boldsymbol{\rho}_q(T)
\right].
\label{eqDistEnt}
\end{align}

\begin{figure}[!htb]
    \centering
    \includegraphics[scale=0.4,keepaspectratio]{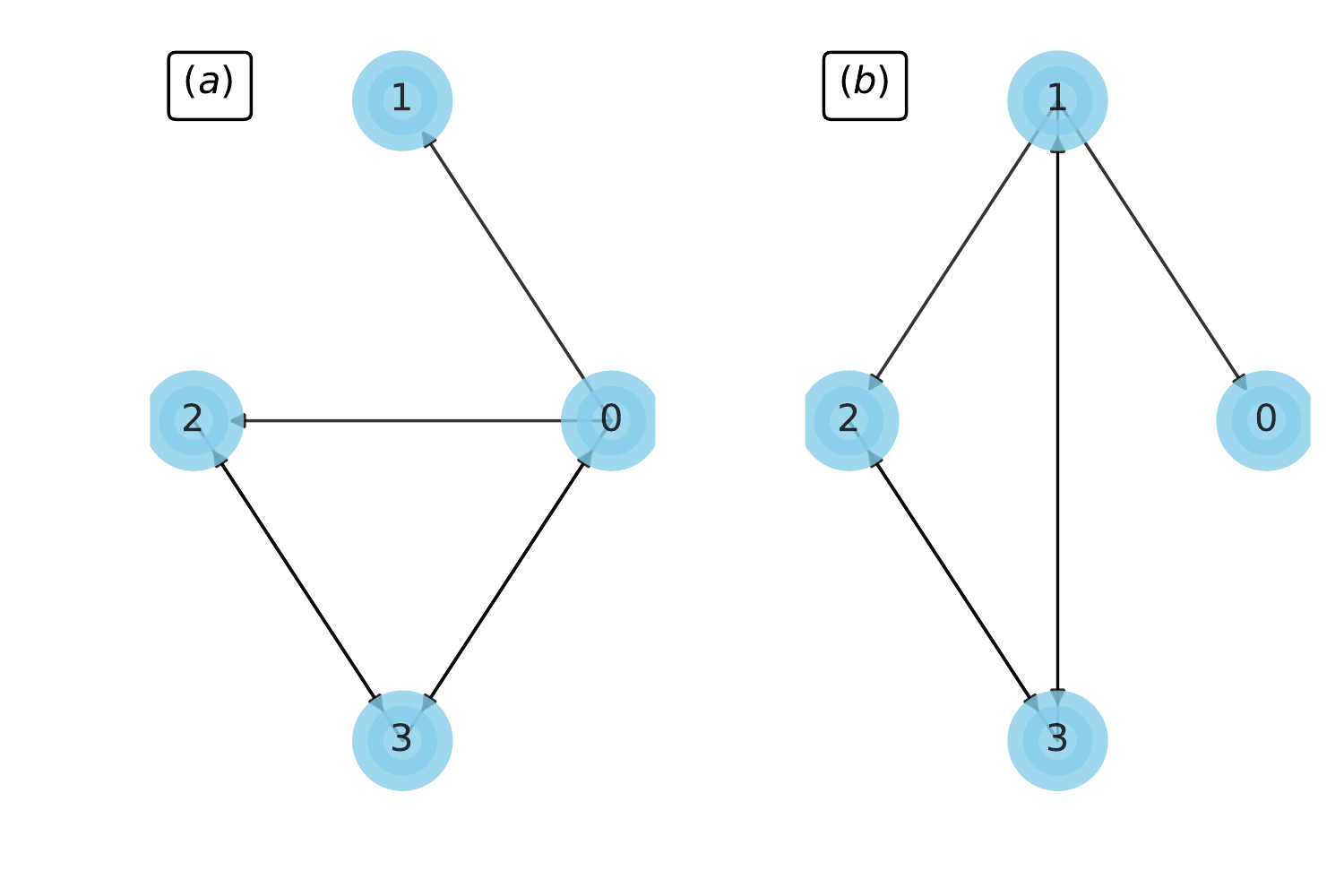}
    \caption{The networks in (a) and (b) are isomorphic in the sense that they can be mapped one into the other by changing the indexes $0\to 1, 1\to 0$.}
    \label{figSameGraph}
\end{figure}

However, as can be noted the term within the trace depends on the product of distinct matrices. Thus, even if two networks  presented in \figref{figSameGraph} are isomorphic, the measure of entropic dissimilarity is nonzero when it desirable be null.	
Another issue related with the entropic similarity approach is that measure  cannot be used to  compare networks with different number of nodes.  This is interesting also because the largest weakly connected component generated by a model does not necessarily have the same size as the overall number of nodes. At the same time, such measure has a high computational cost.

In this work, we suggest the application of the kernel polynomial method jointly with the Wasserstein metric in order to quantify the dissimilarity between the  directed networks.  It should be emphasized that this combination of approaches is only possible given that the the magnetic Laplacian is a Hermitian operator.

\subsection{Comparasion of large directed networks: The KPM method and the Wasserstein Metric}

In order to compute the spectral distance between two networks it is necessary to compute the spectral density,

\begin{align}
    \rho_q(\lambda) = 
        \frac{1}{|V|}
        \sum\limits_{l=1}^{|V|}
            \delta(\lambda-\lambda_{l,q})
            \label{eq_spectral_density}
\end{align}
which has complexity order $O(|V|^3)$. As such, this approach becomes unfeasible for larger networks ($|V|>10^5$).  Fortunately,  the magnetic Laplacian matrix is Hermitian and often sparse, so that the method known as kernel polynomial (KPM) can be considered~\cite{KernelPolynomial2006} for estimating the $\rho_q$.

The KPM objective consists in calculating a simplex $\{
		\mathbf p \in \mathbb{R}_+^{n}
        : \sum\limits_{i=1}^n p_i = 1
    \}$ which allows to define a discrete measure
    \begin{align}
    \alpha_q = \sum\limits_{i=1}^n p_{i,q} \delta_{\lambda_{i,q}}
    \label{eq_discrete_measure}
    \end{align} that approximates the ~\eqreff{eq_spectral_density} with enough accuracy.
    
	This method is based in two approximations.  The first is that any continuous real  function   in an interval $[-1, 1]$ can be expanded in terms of Chebyshev polynomials~\cite{Chebyshev2001}, allowing the spectral density to be approximated by $n$ terms.  The second approximation consists in evaluating the traces associated with the terms of that expansion using Hutchinson's approach~\cite{Hutchinson1990}. In essence, the trace of a sparse matrix function can be aproximated by the product of this function by a set of random vectors.  The oscilations induced by these approximations can be smoothed by subsequently applying a known kernel, which in the present work corresponds to the Jackson kernel~\cite{Jackson1912}.  Therefore, the KPM allows the spectrum of magnetic Laplacian  to be estimated by using algorithms with near-linear computational cost.  In this way, it becomes possible to estimate the spectral density and measurements such as entropy and specific heat even in the case of very large networks containing million of nodes.    
 
Given that it is possible to effectively estimate the magnetic Laplacian spectral density, we can employ Wasserstein metric in order to define distances between directed networks. 
 
For instance, let the set of admissible couplings  of two probability distributions $\alpha_q$ and $\tilde \alpha_q$, given as

 \begin{align}
 U(\alpha, \tilde \alpha)=
 	\{
    	\mathbf U \in \mathbb{R}^{|V|\times|\tilde V|}_{+}
        :
        \mathbf U \mathbf 1_{|\tilde V|}=\mathbf p, \  \ 
        \mathbf U^{T} \mathbf 1_{|V|}=\mathbf {\tilde p}
    \}
 \end{align}
 For a $d \ge 1$  
 a d-Wasserstein distance between the two measures is given by
 
 \begin{align}
 W_d(p, \tilde p, q) = 
 	\left(
    \min_{P \in U(\alpha, \tilde \alpha)}
 	\left[
    \sum\limits_{i,j}	|\lambda_{i,q} - \tilde \lambda_{j,q}| ^d U_{i,j}
 	\right]
    \right)^{1/d}.
 \end{align}
This function has several desired characteristics, such as: it is a metric, it can be applied to networks with different number of nodes, it has relaxed  implementations that allow the distance value to be obtained with a smaller computational cost.
 
	The task of estimating the value of a parameter used to generate a given network, such as the connecting probability in the ER model, can be approached by seeking for a minimum Wasserstein distance between the original network and a set of $n_{exp}$ networks synthesized by considering several parameters.  In this work, we chose a set of $n_q$ charges from which the magnetic Laplacians of each candidate network is obtained, then KPM is used to obtain the respective spectra, and the minimal distance between the latter and the original is determined by using the Wasserstein distance
 
    \begin{align}
    \langle W_d\rangle(p)
    =
    \frac{1}{n_{exp}}
    \sum\limits_{ p\in  P}
        \left(
        	\frac{1}{n_q}
            \sum\limits_{q\in Q}
            	W_d(p, \tilde p, q)
        \right).
        \label{eqWdMean}
\end{align}

\section{Results}

\subsection{Community structures in network and  spectral symmetries}\label{secSupMatStructureAndCv}
\label{sectionAnaliseDeFluxo}

As a first step to address the problem of characterizing directed complex networks by using the magnetic Laplacian formalism, we derive some analytic and numerical results relating network structure and the spectrum of the magnetic Laplacian operator.

First, we aim at studying the influence of community  structure in  directed networks on the magnetic Laplacian spectrum and, consequently, on the specific heat, $c_\lambda$. We assume that the connections within the communities, $\mathbf{W_{in}}$, as well as between the communities, $\mathbf{W_{out}}$, are not differentiated between the structures.  Under this hypothesis, the adjacency matrix can be organized as follows, assuming $N_f$ communities (henceforth, we take $N_f>2$):

\begin{eqnarray}
\mathbf{W} = \begin{bmatrix}
\mathbf{W_{in}}& \mathbf{W_{out}}  & \mathbf{0}_{N_c}& \dots &\mathbf{0}_{N_c}\\ 
\mathbf{0}_{N_c}& \mathbf{W_{in}} &  \mathbf{W_{out}}& \dots&\mathbf{0}_{N_c} \\ 
\vdots & \vdots &  \vdots& \ddots&\vdots \\ 
\mathbf{W_{out}}&  \mathbf{0}_{N_c}&   \mathbf{0}_{N_c}&\dots&\mathbf{W_{in}}
\end{bmatrix},
\end{eqnarray}
where $\mathbf 0_{N_c}$ is a null matrix $N_c\times N_c$.  For generality's sake $\mathbf{W_{in}}$ and $\mathbf{W_{out}}$ can be constructed in arbitrary form.

The magnetic Laplacian expressed as discussed above has the following organization:

\begin{eqnarray}
\mathbf{H}_q = \begin{bmatrix}
\mathbf{H_{in}}& \mathbf{H_{out}}   & \mathbf{0}_{N_c}& \dots &\mathbf{H_{out}}^\dagger\\ 
\mathbf{H_{out}}^\dagger & \mathbf{H_{in}} &   \mathbf{H_{out}}  & \dots&\mathbf{0}_{N_c} \\ 
\vdots & \vdots &  \vdots& \ddots&\vdots \\ 
 \mathbf{H_{out}}  &  \mathbf{0}_{N_c}&   \mathbf{0}_{N_c}&\dots&\mathbf{H_{in}}
\end{bmatrix},
\label{eqHamilOrig}
\end{eqnarray}
note that this matrix is circulant, i.e.
\begin{eqnarray}
\mathbf{H}_q = \begin{bmatrix}
\mathbf{h}_0& \mathbf{h}_1    & \dots &\mathbf{h}_{N_f - 1}\\ 
\mathbf{h}_{N_f - 1}& \mathbf{h}_{0} &    \dots&\mathbf{h}_{N_f - 2} \\ 
\vdots & \vdots &  \ddots&\vdots \\ 
\mathbf{h}_1   &  \mathbf{h}_{2} &\dots&\mathbf{h}_0
\end{bmatrix}.
\end{eqnarray}
	Observe that $\mathbf{H}_q$ is a specific case of a Toepltiz matrix \cite{gray2006toeplitz}, so that the eigenvalues  can be obtained  considering the property that all the columns in the original matrix can be expressed as cyclic permutations of the first column.

Our objective now is to find the set $\{\lambda_u\}$ such that  
\begin{eqnarray}
\mathbf{H}_q |\psi_u\rangle = \lambda_u |\psi_u\rangle.
\label{eqAutoVec}
\end{eqnarray}

As known from literature~\cite{gray2006toeplitz}, the
eigenvectors of a cyclic matrix can be obtained as
\begin{align}
|\psi_u\rangle= \begin{bmatrix}
 |\phi\rangle\\ 
\rho_u |\phi\rangle \\ 
\vdots \\ 
\rho_u^{N_f -1} |\phi\rangle
\end{bmatrix},
\label{eqEigenVec}
\end{align}
where  $u \in \{0,\dots, N_f-1\}$ and $\rho_u=\rho_{N_f-u}^\star= \exp(\frac{2\pi i u}{N_f})$.
Substituting this eigenvector
\eqreff{eqEigenVec} into \eqreff{eqAutoVec}, allows the block equation induced by the first row to be solved as
\begin{align}
\mathbf{\tilde H}_u|\psi_u\rangle=\sum\limits_{l=0}^{N_f-1}\mathbf{h}_l\rho_{l\cdot u} |\psi_u\rangle = \lambda_u |\psi_u\rangle,
\end{align}

The above equation can be simplified  introducing the variable
\begin{eqnarray}
m_f =\begin{cases}
\frac{N_f + 1}{2} \text{ if } N_f \text{ is odd},\\
\frac{N_f}{2} \text{ if } N_f \text{ is even}
\end{cases},
\end{eqnarray}
 and by taking into account that $\mathbf{H_N}$ is Hermitian consequently $\mathbf h_j = \mathbf h_{N_f - j}^\dagger$.
 
The simplified version is given as
\begin{align}
\mathbf{\tilde H}_u=\mathbf{h}_0
+\sum\limits_{l=1}^{m_f-1}
    \left(
    \mathbf{h}_l\rho_{l\cdot u} +
    \mathbf{h}_l^\dagger\rho_{l\cdot u}^\star
    \right)+ \mathbf{\Delta},
\end{align}
where
\begin{align}
\mathbf{\Delta}= \begin{cases}
 \mathbf{0}_{N_c}\text{ if } N_f \text{ is odd},\\
 (-1)^u\mathbf{h}_{m_f} \text{ if } N_f \text{ is even}
 \end{cases}.
\end{align}

Since in the flow structure $\mathbf{\Delta}=\mathbf{0}_{N_c}$, and only three instances $\mathbf{h}_{u}$ are non-null, we have
\begin{align}
\mathbf{\tilde H}_u=\mathbf{h}_0
+
\mathbf{h}_1\rho_{u} +
\mathbf{h}_1^\dagger\rho_{u}^\star,
\end{align}
Replacing the operators $\mathbf{h}$ by their respective counterparts in equation \eqreff{eqHamilOrig}, we obtain the following expression for the $u$-th matrix in a network with $N_f$ blocks, 
\begin{align}
\mathbf{\tilde H}_u=\mathbf{H_{in}}
+
e^{\frac{2\pi i u}{N_f}}\mathbf{H_{out}} +
e^{-\frac{2\pi i u}{N_f}}\mathbf{H_{out}}^\dagger.
\label{eqFluxGeral}
\end{align}
In the following sections we will investigate how distinct $\mathbf{H_{in}}$ influence $c_\lambda$.

\subsubsection{Uniform Connections}
\begin{figure}[!htb]
    \centering
    \includegraphics[width=\columnwidth,keepaspectratio]{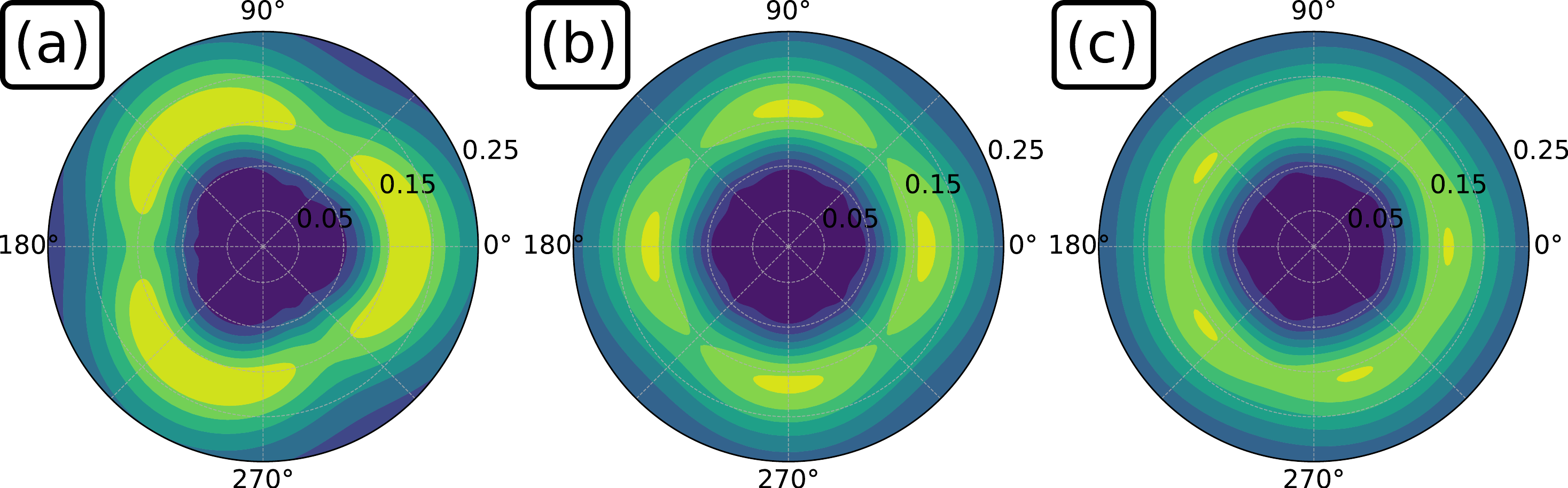}
   \caption{Specific heat (shown in colors) in terms of the charge $2\pi q$ (polar coordinates) and temperature (radial coordinate) for $N_f=3$(a), $4$(b) and $5$(c), assuming $N_c=45$. This plot was derived from Equation~\eqreff{eqLambdaFinal}.  }
   \label{figNf3and4and5}
 \end{figure}
Uniform connection is characterized by having the degree of each vertex given as
$[\mathbf D_{ii}] = d = 2N_c - 1$.
Consequently, the intrablock of the magnetic Laplacian is  
\begin{eqnarray}
\mathbf{H_{in}} =\frac{\mathbf{I}_{N_c}(1+d)-\mathbf{1}_{N_c}}{d},
\end{eqnarray}
and the interblock defining the connections between the modular structures is given as
\begin{eqnarray}
\mathbf{H_{out}} = -\frac{\exp(2\pi i q)}{2d}\mathbf{1}_{N_c}.
\end{eqnarray}
Substituting the two previous equations into \eqreff{eqFluxGeral}, 
 $\mathbf{\tilde H}_u$ can be obtained as
\begin{align}
\mathbf{\tilde H}_u&= 
 \frac{\mathbf{I}_{N_c}(1+d)-\mathbf{1}_{N_c}}{d}\nonumber\\
 & -2\frac{\cos(2\pi( \frac{u}{N_f} -q))}{2d}\mathbf{1}_{N_c},
\end{align}
observe that $\mathbf{\tilde H}_u$ is a circulant matrix. Due that   let $v\in\{0,...,N_c-1\}$, and define 
\begin{eqnarray}
m_c =\begin{cases}
\frac{N_c + 1}{2} \text{ if } N_c \text{ is odd},\\
\frac{N_c}{2} \text{ if } N_c \text{ is even}
\end{cases},
\end{eqnarray}
the eigenvalues of  $\mathbf{\tilde H}_u$ can be obtained as

\begin{align}
\lambda_{u,v}=h_0
+\sum\limits_{l=1}^{m_c-1}
\left(
h_l\rho_{l\cdot v} +
h_l^\dagger\rho_{l\cdot v}^\star
\right)+ \Delta.
\label{eqLambdaUniforme}
\end{align}
where
\begin{align}
\Delta= \begin{cases}
0\text{ if } N_c \text{ is odd},\\
(-1)^v h_{m_c} \text{ if } N_c \text{ is even}
\end{cases}.
\end{align}
Replacing $h_l$ by their counterparts in~\eqreff{eqLambdaUniforme} the following eigenvalue
equation can be obtained
\begin{align}
\lambda_{u,v}&=1 -\frac{\cos(2\pi( \frac{u}{N_f} -q))}{d}\nonumber\\
&+\frac{2}{d}\left(
1+\cos(2\pi( \frac{u}{N_f} -q))
\right)f(v, N_c, m_c)
+ \Delta,
\label{eqLambdaFinal}
\end{align}
where $f(v, N_c, m_c) = \sum\limits_{l=1}^{m_c-1}
\cos(\frac{2\pi v l}{N_c})$, such that
\begin{align}
f(v, N_c, m_c)= \begin{cases}
m_c\text{ if } v=0,\\
\frac{\sin(\frac{\pi v m_c}{N_c}) }{\sin(\frac{\pi v }{N_c})}\cos(\frac{\pi v }{N_c}(m_c-1)) \text{ otherwise}
\end{cases}.
\end{align}

The~\eqreff{eqLambdaFinal} indicates a rotation symmetry related to the charge parameter in the modular directed network. 
These symmetries also reflect the behavior of the specific heat petal structure shown in \figref{figNf3and4and5}.

\subsubsection{ Asymmetries in the specific heat petal structures}

\begin{figure}[!htb]
    \centering
    \includegraphics[width=\columnwidth,keepaspectratio]{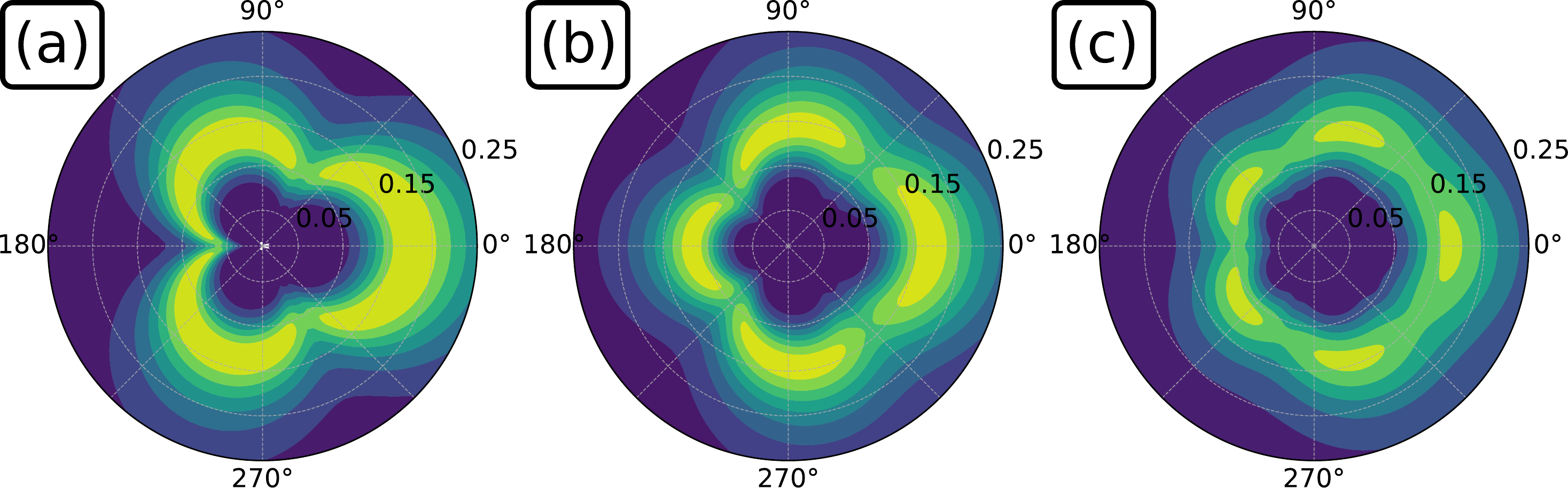}
    \caption{ Specific heat (colors) in terms of the charge 
     $2\pi q$ (angle) and temperature (radius), for $N_f=3$(a), $4$(b) and $5$(c), assuming $N_c=45$. The networks were
generated randomly, imposing the probability of having a
directed edge as $p_c=30\%$. Observe the obtained
asymmetric petals contrasting with the results obtained
previously for the uniform connections.    }
    \label{figRandomNf3and4and5}

\end{figure}

	The results obtained in the previous section helps to understand  the relationship between the modular structures and the magnetic Laplacian spectrum, as well as the specific heat symmetry.  However, these results assume that the inner structures $\mathbf{H_{in}}$ are undirected.  The effect of directionality can be inferred by generating random directions inside the intrablocks, i.e.~ by imposing that $[\mathbf{W_{in}}]_{u,v}$ has probability $p_c < 100\%$ to take value $1$.  	
Adopting $p_c=30\%$, we calculate the specific heat by using numeric diagonalization, yielding the structures in \figref{figRandomNf3and4and5}.  We can observe the obtained petals are not symmetric, unlike what had been observed for uniform connections.

\subsection{Model characterization of directed graphs}

	In this section we address the task of characterization of distinct networks models through the spectra of magnetic Laplacian.  In particular, given a set of measurements obtained from a graph, can we infer which model created that graph?   In this work, we opted to use the specific heat, \cv, as a feature of measurement of graphs, in order to address the  question above.  	
As shown in \figref{figNets2Cv}, the  \cv \ \ measures yielded specific behavior for different  models, therefore providing valuable information that can be use to identify and discriminate between different complex networks models.

In order to evaluate the efficiency of using \cv \ \ as a fingerprint of a directed network, we built a dataset with $2000$ network samples with types Erdős–Rényi (ER),  Barabási (BA), Bollobás's et al scale-free model~\cite{Bollobas2003} (SF), Watts-Strogatz  (WS),  and SBM  with $3$ and $4$ blocks.

\begin{figure}[!htb]
    \centering    \includegraphics[width=\columnwidth,keepaspectratio]{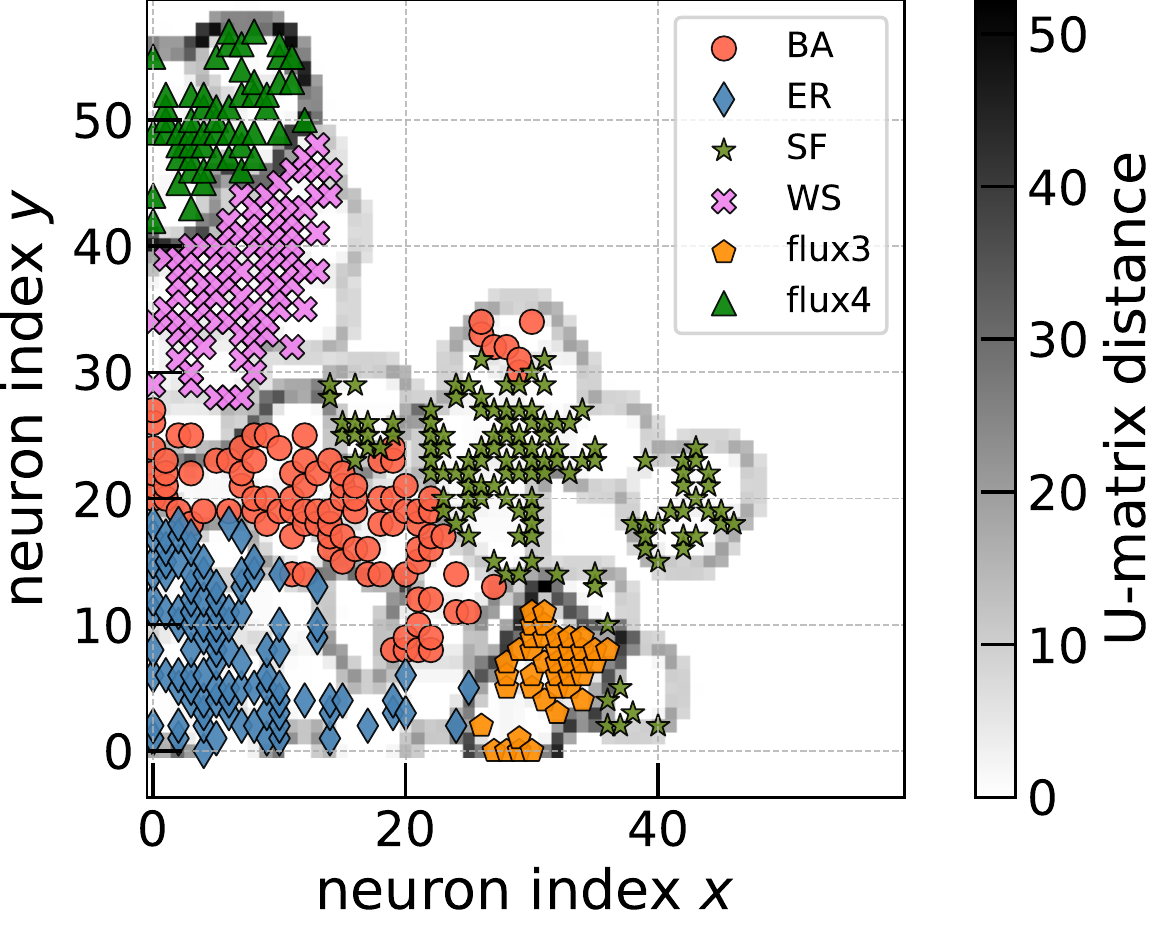}
    \caption{SOM mapping of six types of complex networks represented by the specific heat approach.  The \emph{neuron index x} and \emph{neuron index y}  correspond to neurons in the SOM cortical space.  The distances between neighboring neurons (U-matrix) are indicated in gray.  A good separation between the types of networks can be observed.}
 	\label{figSOMLabels}
\end{figure}

Then, self organizing maps (SOMs), namely a method for non-supervised clustering~\cite{SOM2019PRB}, were trained with the obtained \cv ~values and the obtained regions were subsequently labeled.  This was done by feeding each training data into the SOM and choosing the neuron that exhibited highest activation.  As indicated by the results  shown in  \figref{figSOMLabels}, networks belonging to the same class have been mapped into nearby neurons, defining respective clusters.  So, the SOM was able, without previous knowledge  to find the patterns of \cv \ \ associated to the considered types of networks. 

From what we have seen, we can conclude  that the suggested magnetic Laplacian approach is able, at least for the considered cases, to properly characterize the model of given networks.
For this reason, in a similar manner to that which has been applied in condensed matter physics, ``SOM'' proved to be a powerful technique for characterizing complex networks when we see these networks through the lens of statistical mechanics and magnetic Laplacians.

\begin{figure}[!htb]
    \centering
  \includegraphics[width=\columnwidth,keepaspectratio]{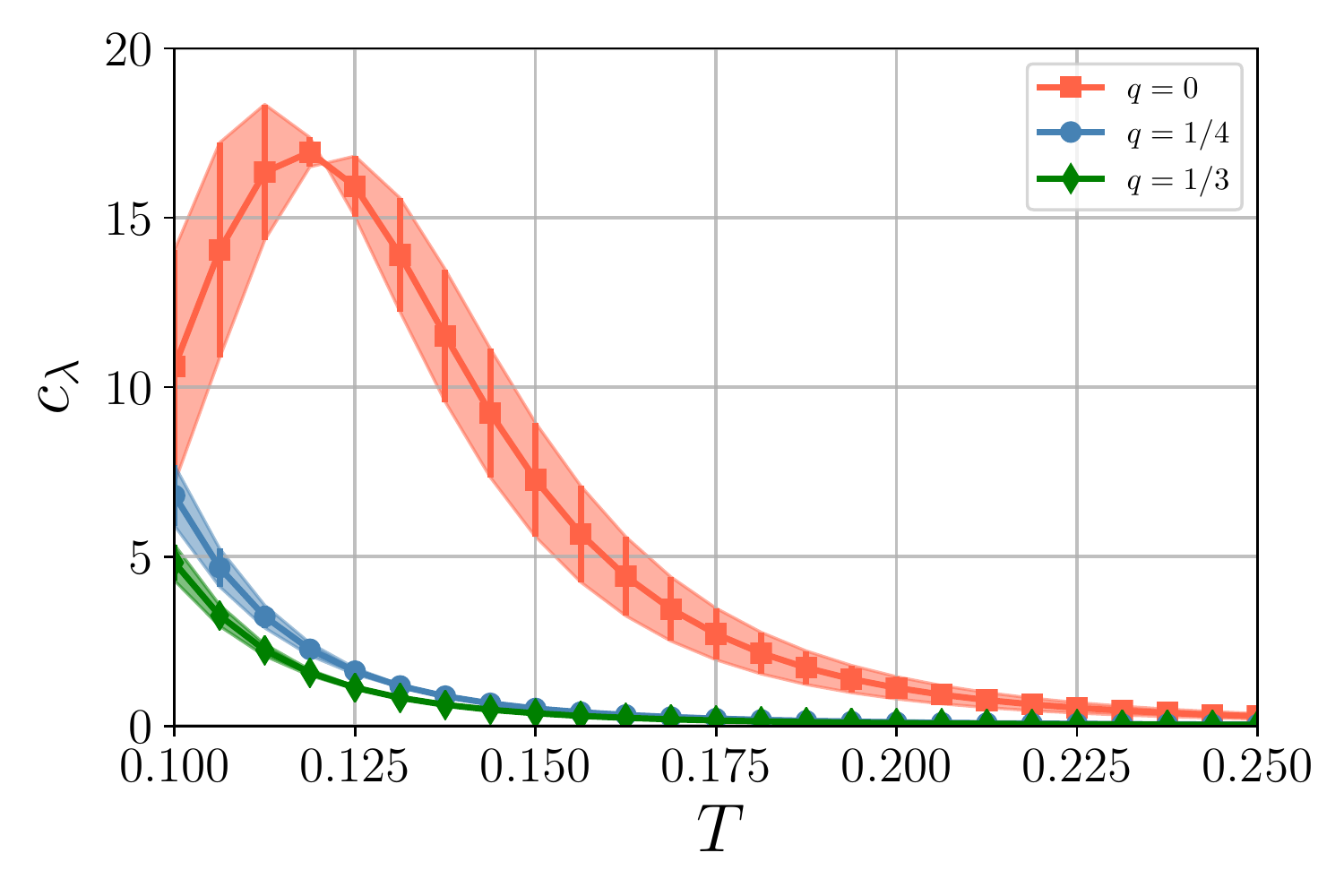}
    \caption{Approximated specific heat for a network with  
$|V|=3000$,  $N_f=3$, $p_c=0.25$ and $p_d=0.5$. In the application of KPM method the expansion was truncated at  $40$ first terms and the stochastic trace approximation used $25$ random vectors. The error-bars represent the deviation between the exact value (obtained numerically) and the approximated value calculated by the KPM method and using numerical integration. }
      \label{figCvKPM}
\end{figure}

	Given that many real-world networks contain a large number of nodes, a question arises regarding the feasibility using spectral quantities for their characterization.  As described in the methodology section, thanks to the magnetic Laplacian formalism, KPM can be used as a means to estimate spectral density measurements.  For instance, given a modular directed network  we obtained the exact  and KPM-approximated values of the specific heat for different temperatures and charge values.  The approximated specific heat is shown in~\figref{figCvKPM}.  The error bars indicate a  small dispersion, corroborating the potential of the KPM approach for studying the spectral properties of complex networks.

\subsection{Directed network parameter Inference}

The results shown in  \figref{figSOMLabels}  indicates that, given a network $\tilde G$, we can infer which model was responsible for generating it. In addition, to complete the task of characterizing a network it is necessary to find the network which most closely resembles $\tilde G$ among several networks created with distinct parameters while fixing the model.  

	In this section, we explore the problem of inferring the parameters  of  models using the  spectra of  the magnetic Laplacian..

\begin{figure}[!htb]
    \centering
  \includegraphics[width=\columnwidth,keepaspectratio]{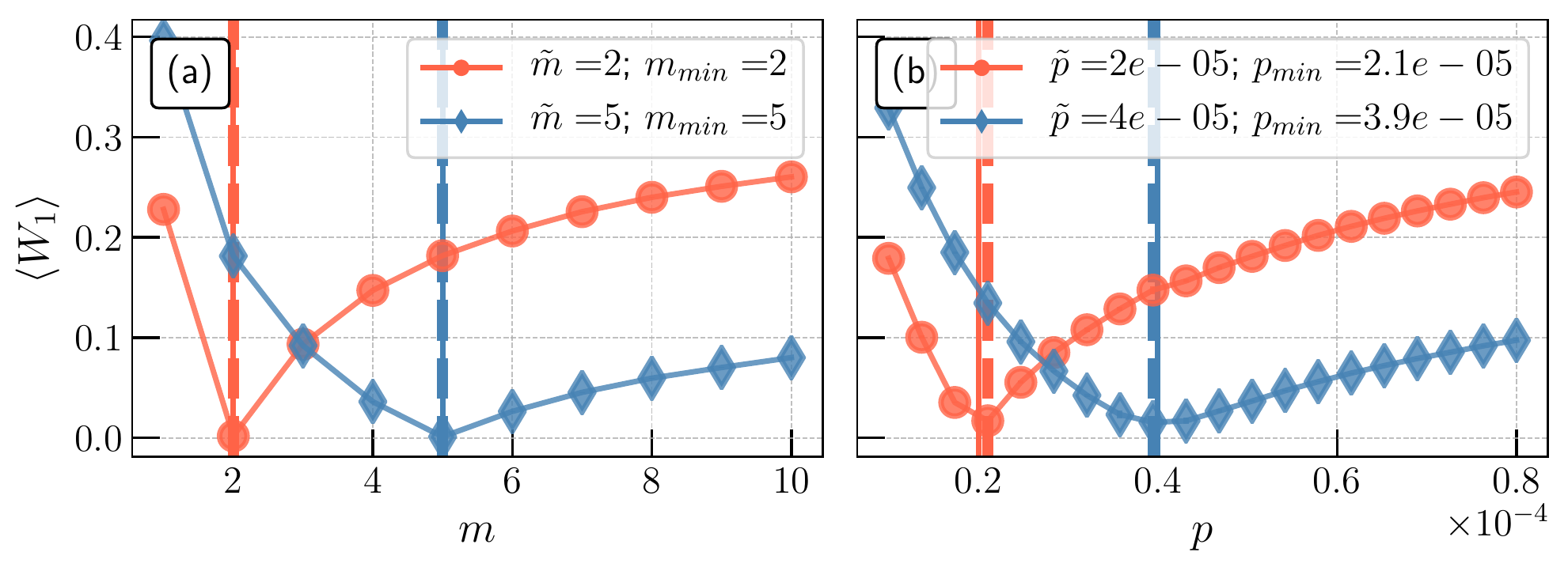}
    \caption{The curves in (a) and (b) represent the  mean of 1-Wasserstein  distances \eqreff{eqWdMean}, respectively to BA and ER, in terms of the parameters adopted for network generation, considering
 $N_{exp}=5$, $|V|=10^5$, and $Q=\{0, 1/3\}$. For spectral estimation using the KPM  was used $100$ terms of expansion and $20$ random vectors.}
      \label{figDistKPMWs}
\end{figure}
In order to argue that the Wasserstein metric can be used combined with the KPM approach as a means to estimate the network model parameters with sufficient precision,  we  study the problem of infering  the conecting probabilities $\tilde p$ of ER networks and the out-degree $\tilde m$ of BA networks, both with approximately $10^5$ nodes.

	In~\figref{figDistKPMWs} the continuous vertical lines show the	
correct value of the parameter and the vertical dashed	
lines identify the position of the minimal of Eq.\eqref{eqWdMean}, which is the inferred value of the parameter. 	
By using KPM with the  $100$ first terms of the Chebyshev polynomial and approximating the trace by using 	
$20$ random vectors, we observe  the parameters can be inferred with good accuracy.

\section{Conclusions}

Directed networks can be used to represent several real-world structures and problems.  As a consequence, several approaches have been proposed aimed at characterizing and comparing directed networks.  Among these approaches, spectral methods present some particularly interesting properties, such as bearing a direct relationship with the structural and dynamical aspects of given networks.  However, when applied to directed networks, the usual Laplacian operator yields complex eigenvalues, which are difficult to treat and interpret.  Nevertheless, the hermiticity property of the magnetic Laplacian allows a set of real eigenvalues to be associated with a weighted directed network.  We showed here that real eigenvalues and the associated charge parameter convey information about the network, more specifically regarding its  mesoscale structures  and the spectral and specific heat symmetry.

In order to extend the proposed methodology to larger networks containing hundreds of thousands of nodes, we showed  the KPM method can be combined with the magnetic Laplacian approach. This combination allowed to estimate the spectral density of the magnetic operator with remarkable efficiency and accuracy. Given that we could estimate the spectral density of the magnetic Laplacian, we showed that the study of spectral geometry under the Wasserstein metric can be used as a tool to infer parameters of networks with low relative errors.

The reported contributions pave the way to a number of future developments and applications involving directed complex networks.   For instance, these methods can be  applied to study several other theoretical and real world structures, including fake news dissemination, metabolic networks, neuronal systems, to name but a few possibilities.  It would also be interesting to perform studies using random matrix theory in order to infer relationships between topology and spectra for more general complex networks.  Since we deal only with spectral information, the results presented in this paper could also be immediately applied to multiplex networks.

\section*{Acknowledgements}
The authors thank Thomas Peron, Henrique F. de Arruda, Paulo E. P. Burke and  Filipi N. Silva  for all suggestions and  useful discussions. Bruno Messias thanks CAPES for financial support. Luciano da F. Costa thanks CNPq (grant no. 307085/2018-0) and NAP-PRP-USP for sponsorship. This work has been supported also by FAPESP
grant 15/22308-2. Research carried out using the computational resources of the Center for Mathematical Sciences Applied to Industry (CeMEAI) funded by FAPESP (grant 2013/07375-0).

\section*{DATA AVAILABLITY }
Data sharing is not applicable to this article as no new data were created or analyzed in this
study. However, our implementation of KPM method  it is available at \href{https://github.com/stdogpkg/emate}{github.com/stdogpkg/emate}. In addition eMaTe also allows to estimate trace functions of symmetric adjacency matrices with a good accuracy and computational efficiency. 

\bibliography{biblio.bib}

\begin{thebibliography}{31}%
\makeatletter
\providecommand \@ifxundefined [1]{%
 \@ifx{#1\undefined}
}%
\providecommand \@ifnum [1]{%
 \ifnum #1\expandafter \@firstoftwo
 \else \expandafter \@secondoftwo
 \fi
}%
\providecommand \@ifx [1]{%
 \ifx #1\expandafter \@firstoftwo
 \else \expandafter \@secondoftwo
 \fi
}%
\providecommand \natexlab [1]{#1}%
\providecommand \enquote  [1]{``#1''}%
\providecommand \bibnamefont  [1]{#1}%
\providecommand \bibfnamefont [1]{#1}%
\providecommand \citenamefont [1]{#1}%
\providecommand \href@noop [0]{\@secondoftwo}%
\providecommand \href [0]{\begingroup \@sanitize@url \@href}%
\providecommand \@href[1]{\@@startlink{#1}\@@href}%
\providecommand \@@href[1]{\endgroup#1\@@endlink}%
\providecommand \@sanitize@url [0]{\catcode `\\12\catcode `\$12\catcode
  `\&12\catcode `\#12\catcode `\^12\catcode `\_12\catcode `\%12\relax}%
\providecommand \@@startlink[1]{}%
\providecommand \@@endlink[0]{}%
\providecommand \url  [0]{\begingroup\@sanitize@url \@url }%
\providecommand \@url [1]{\endgroup\@href {#1}{\urlprefix }}%
\providecommand \urlprefix  [0]{URL }%
\providecommand \Eprint [0]{\href }%
\providecommand \doibase [0]{http://dx.doi.org/}%
\providecommand \selectlanguage [0]{\@gobble}%
\providecommand \bibinfo  [0]{\@secondoftwo}%
\providecommand \bibfield  [0]{\@secondoftwo}%
\providecommand \translation [1]{[#1]}%
\providecommand \BibitemOpen [0]{}%
\providecommand \bibitemStop [0]{}%
\providecommand \bibitemNoStop [0]{.\EOS\space}%
\providecommand \EOS [0]{\spacefactor3000\relax}%
\providecommand \BibitemShut  [1]{\csname bibitem#1\endcsname}%
\let\auto@bib@innerbib\@empty
\bibitem [{\citenamefont {Kac}(1966)}]{Kac1966}%
  \BibitemOpen
  \bibfield  {author} {\bibinfo {author} {\bibfnamefont {M.}~\bibnamefont
  {Kac}},\ }\bibfield  {title} {\enquote {\bibinfo {title} {Can one hear the
  shape of a drum?}}\ }\href {\doibase 10.1080/00029890.1966.11970915}
  {\bibfield  {journal} {\bibinfo  {journal} {The American Mathematical
  Monthly}\ }\textbf {\bibinfo {volume} {73}},\ \bibinfo {pages} {1--23}
  (\bibinfo {year} {1966})}\BibitemShut {NoStop}%
\bibitem [{\citenamefont {Giraud}\ and\ \citenamefont
  {Thas}(2010)}]{RevModPhysShapeDrum2010}%
  \BibitemOpen
  \bibfield  {author} {\bibinfo {author} {\bibfnamefont {O.}~\bibnamefont
  {Giraud}}\ and\ \bibinfo {author} {\bibfnamefont {K.}~\bibnamefont {Thas}},\
  }\bibfield  {title} {\enquote {\bibinfo {title} {Hearing shapes of drums:
  Mathematical and physical aspects of isospectrality},}\ }\href {\doibase
  10.1103/RevModPhys.82.2213} {\bibfield  {journal} {\bibinfo  {journal} {Rev.
  Mod. Phys.}\ }\textbf {\bibinfo {volume} {82}},\ \bibinfo {pages}
  {2213--2255} (\bibinfo {year} {2010})}\BibitemShut {NoStop}%
\bibitem [{\citenamefont {Aasen}, \citenamefont {Bhamre},\ and\ \citenamefont
  {Kempf}(2013)}]{Aasen2013}%
  \BibitemOpen
  \bibfield  {author} {\bibinfo {author} {\bibfnamefont {D.}~\bibnamefont
  {Aasen}}, \bibinfo {author} {\bibfnamefont {T.}~\bibnamefont {Bhamre}}, \
  and\ \bibinfo {author} {\bibfnamefont {A.}~\bibnamefont {Kempf}},\ }\bibfield
   {title} {\enquote {\bibinfo {title} {Shape from sound: Toward new tools for
  quantum gravity},}\ }\href {\doibase 10.1103/physrevlett.110.121301}
  {\bibfield  {journal} {\bibinfo  {journal} {Physical Review Letters}\
  }\textbf {\bibinfo {volume} {110}} (\bibinfo {year} {2013}),\
  10.1103/physrevlett.110.121301}\BibitemShut {NoStop}%
\bibitem [{\citenamefont {Cosmo}\ \emph {et~al.}(2018)\citenamefont {Cosmo},
  \citenamefont {Panine}, \citenamefont {Rampini}, \citenamefont {Ovsjanikov},
  \citenamefont {Bronstein},\ and\ \citenamefont
  {Rodolà}}]{isospectralization2018}%
  \BibitemOpen
  \bibfield  {author} {\bibinfo {author} {\bibfnamefont {L.}~\bibnamefont
  {Cosmo}}, \bibinfo {author} {\bibfnamefont {M.}~\bibnamefont {Panine}},
  \bibinfo {author} {\bibfnamefont {A.}~\bibnamefont {Rampini}}, \bibinfo
  {author} {\bibfnamefont {M.}~\bibnamefont {Ovsjanikov}}, \bibinfo {author}
  {\bibfnamefont {M.~M.}\ \bibnamefont {Bronstein}}, \ and\ \bibinfo {author}
  {\bibfnamefont {E.}~\bibnamefont {Rodolà}},\ }\href
  {https://arxiv.org/pdf/1811.11465.pdf} {\enquote {\bibinfo {title}
  {Isospectralization, or how to hear shape, style, and correspondence},}\ }
  (\bibinfo {year} {2018})\BibitemShut {NoStop}%
\bibitem [{\citenamefont {Cvetkovi{\'c}}(1971)}]{cvetkovic1971graphs}%
  \BibitemOpen
  \bibfield  {author} {\bibinfo {author} {\bibfnamefont {D.~M.}\ \bibnamefont
  {Cvetkovi{\'c}}},\ }\bibfield  {title} {\enquote {\bibinfo {title} {Graphs
  and their spectra},}\ }\href@noop {} {\bibfield  {journal} {\bibinfo
  {journal} {Publikacije Elektrotehni{\v{c}}kog fakulteta. Serija Matematika i
  fizika}\ ,\ \bibinfo {pages} {1--50}} (\bibinfo {year} {1971})}\BibitemShut
  {NoStop}%
\bibitem [{\citenamefont {van Dam}\ and\ \citenamefont
  {Haemers}(2003)}]{vanDam2003}%
  \BibitemOpen
  \bibfield  {author} {\bibinfo {author} {\bibfnamefont {E.~R.}\ \bibnamefont
  {van Dam}}\ and\ \bibinfo {author} {\bibfnamefont {W.~H.}\ \bibnamefont
  {Haemers}},\ }\bibfield  {title} {\enquote {\bibinfo {title} {Which graphs
  are determined by their spectrum?}}\ }\href {\doibase
  10.1016/s0024-3795(03)00483-x} {\bibfield  {journal} {\bibinfo  {journal}
  {Linear Algebra and its Applications}\ }\textbf {\bibinfo {volume} {373}},\
  \bibinfo {pages} {241--272} (\bibinfo {year} {2003})}\BibitemShut {NoStop}%
\bibitem [{\citenamefont {Sarkar}\ and\ \citenamefont
  {Jalan}(2018)}]{Sarkar2018}%
  \BibitemOpen
  \bibfield  {author} {\bibinfo {author} {\bibfnamefont {C.}~\bibnamefont
  {Sarkar}}\ and\ \bibinfo {author} {\bibfnamefont {S.}~\bibnamefont {Jalan}},\
  }\bibfield  {title} {\enquote {\bibinfo {title} {Spectral properties of
  complex networks},}\ }\href {\doibase 10.1063/1.5040897} {\bibfield
  {journal} {\bibinfo  {journal} {Chaos: An Interdisciplinary Journal of
  Nonlinear Science}\ }\textbf {\bibinfo {volume} {28}},\ \bibinfo {pages}
  {102101} (\bibinfo {year} {2018})}\BibitemShut {NoStop}%
\bibitem [{\citenamefont {Wang}, \citenamefont {Wilson},\ and\ \citenamefont
  {Hancock}(2017)}]{Wang2017}%
  \BibitemOpen
  \bibfield  {author} {\bibinfo {author} {\bibfnamefont {J.}~\bibnamefont
  {Wang}}, \bibinfo {author} {\bibfnamefont {R.~C.}\ \bibnamefont {Wilson}}, \
  and\ \bibinfo {author} {\bibfnamefont {E.~R.}\ \bibnamefont {Hancock}},\
  }\bibfield  {title} {\enquote {\bibinfo {title} {Detecting alzheimer's
  disease using directed graphs},}\ }in\ \href {\doibase
  10.1007/978-3-319-58961-9_9} {\emph {\bibinfo {booktitle} {Graph-Based
  Representations in Pattern Recognition}}}\ (\bibinfo  {publisher} {Springer
  International Publishing},\ \bibinfo {year} {2017})\ pp.\ \bibinfo {pages}
  {94--104}\BibitemShut {NoStop}%
\bibitem [{\citenamefont {Anand}\ and\ \citenamefont
  {Bianconi}(2009)}]{Ginestra2009}%
  \BibitemOpen
  \bibfield  {author} {\bibinfo {author} {\bibfnamefont {K.}~\bibnamefont
  {Anand}}\ and\ \bibinfo {author} {\bibfnamefont {G.}~\bibnamefont
  {Bianconi}},\ }\bibfield  {title} {\enquote {\bibinfo {title} {Entropy
  measures for networks: Toward an information theory of complex topologies},}\
  }\href {\doibase 10.1103/PhysRevE.80.045102} {\bibfield  {journal} {\bibinfo
  {journal} {Phys. Rev. E}\ }\textbf {\bibinfo {volume} {80}},\ \bibinfo
  {pages} {045102} (\bibinfo {year} {2009})}\BibitemShut {NoStop}%
\bibitem [{\citenamefont {Dehmer}\ and\ \citenamefont
  {Mowshowitz}(2011)}]{Dehmer2011}%
  \BibitemOpen
  \bibfield  {author} {\bibinfo {author} {\bibfnamefont {M.}~\bibnamefont
  {Dehmer}}\ and\ \bibinfo {author} {\bibfnamefont {A.}~\bibnamefont
  {Mowshowitz}},\ }\bibfield  {title} {\enquote {\bibinfo {title} {A history of
  graph entropy measures},}\ }\href {\doibase 10.1016/j.ins.2010.08.041}
  {\bibfield  {journal} {\bibinfo  {journal} {Information Sciences}\ }\textbf
  {\bibinfo {volume} {181}},\ \bibinfo {pages} {57--78} (\bibinfo {year}
  {2011})}\BibitemShut {NoStop}%
\bibitem [{\citenamefont {Ye}\ \emph {et~al.}(2015)\citenamefont {Ye},
  \citenamefont {Comin}, \citenamefont {Peron}, \citenamefont {Silva},
  \citenamefont {Rodrigues}, \citenamefont {Costa}, \citenamefont {Torsello},\
  and\ \citenamefont {Hancock}}]{pre2015}%
  \BibitemOpen
  \bibfield  {author} {\bibinfo {author} {\bibfnamefont {C.}~\bibnamefont
  {Ye}}, \bibinfo {author} {\bibfnamefont {C.~H.}\ \bibnamefont {Comin}},
  \bibinfo {author} {\bibfnamefont {T.~K.~D.}\ \bibnamefont {Peron}}, \bibinfo
  {author} {\bibfnamefont {F.~N.}\ \bibnamefont {Silva}}, \bibinfo {author}
  {\bibfnamefont {F.~A.}\ \bibnamefont {Rodrigues}}, \bibinfo {author}
  {\bibfnamefont {L.~d.~F.}\ \bibnamefont {Costa}}, \bibinfo {author}
  {\bibfnamefont {A.}~\bibnamefont {Torsello}}, \ and\ \bibinfo {author}
  {\bibfnamefont {E.~R.}\ \bibnamefont {Hancock}},\ }\bibfield  {title}
  {\enquote {\bibinfo {title} {Thermodynamic characterization of networks using
  graph polynomials},}\ }\href {\doibase 10.1103/PhysRevE.92.032810} {\bibfield
   {journal} {\bibinfo  {journal} {Phys. Rev. E}\ }\textbf {\bibinfo {volume}
  {92}},\ \bibinfo {pages} {032810} (\bibinfo {year} {2015})}\BibitemShut
  {NoStop}%
\bibitem [{\citenamefont {De~Domenico}\ and\ \citenamefont
  {Biamonte}(2016)}]{prx2016}%
  \BibitemOpen
  \bibfield  {author} {\bibinfo {author} {\bibfnamefont {M.}~\bibnamefont
  {De~Domenico}}\ and\ \bibinfo {author} {\bibfnamefont {J.}~\bibnamefont
  {Biamonte}},\ }\bibfield  {title} {\enquote {\bibinfo {title} {Spectral
  entropies as information-theoretic tools for complex network comparison},}\
  }\href {\doibase 10.1103/PhysRevX.6.041062} {\bibfield  {journal} {\bibinfo
  {journal} {Phys. Rev. X}\ }\textbf {\bibinfo {volume} {6}},\ \bibinfo {pages}
  {041062} (\bibinfo {year} {2016})}\BibitemShut {NoStop}%
\bibitem [{\citenamefont {Nicolini}, \citenamefont {Vlasov},\ and\
  \citenamefont {Bifone}(2018)}]{preModelFitEntropy2018}%
  \BibitemOpen
  \bibfield  {author} {\bibinfo {author} {\bibfnamefont {C.}~\bibnamefont
  {Nicolini}}, \bibinfo {author} {\bibfnamefont {V.}~\bibnamefont {Vlasov}}, \
  and\ \bibinfo {author} {\bibfnamefont {A.}~\bibnamefont {Bifone}},\
  }\bibfield  {title} {\enquote {\bibinfo {title} {Thermodynamics of network
  model fitting with spectral entropies},}\ }\href {\doibase
  10.1103/PhysRevE.98.022322} {\bibfield  {journal} {\bibinfo  {journal} {Phys.
  Rev. E}\ }\textbf {\bibinfo {volume} {98}},\ \bibinfo {pages} {022322}
  (\bibinfo {year} {2018})}\BibitemShut {NoStop}%
\bibitem [{\citenamefont {Hart}\ \emph {et~al.}(2015)\citenamefont {Hart},
  \citenamefont {Pade}, \citenamefont {Pereira}, \citenamefont {Murphy},\ and\
  \citenamefont {Roy}}]{PREPereira2015}%
  \BibitemOpen
  \bibfield  {author} {\bibinfo {author} {\bibfnamefont {J.~D.}\ \bibnamefont
  {Hart}}, \bibinfo {author} {\bibfnamefont {J.~P.}\ \bibnamefont {Pade}},
  \bibinfo {author} {\bibfnamefont {T.}~\bibnamefont {Pereira}}, \bibinfo
  {author} {\bibfnamefont {T.~E.}\ \bibnamefont {Murphy}}, \ and\ \bibinfo
  {author} {\bibfnamefont {R.}~\bibnamefont {Roy}},\ }\bibfield  {title}
  {\enquote {\bibinfo {title} {Adding connections can hinder network
  synchronization of time-delayed oscillators},}\ }\href {\doibase
  10.1103/PhysRevE.92.022804} {\bibfield  {journal} {\bibinfo  {journal} {Phys.
  Rev. E}\ }\textbf {\bibinfo {volume} {92}},\ \bibinfo {pages} {022804}
  (\bibinfo {year} {2015})}\BibitemShut {NoStop}%
\bibitem [{\citenamefont {Berkolaiko}(2013)}]{Berkolaiko2013}%
  \BibitemOpen
  \bibfield  {author} {\bibinfo {author} {\bibfnamefont {G.}~\bibnamefont
  {Berkolaiko}},\ }\bibfield  {title} {\enquote {\bibinfo {title} {Nodal count
  of graph eigenfunctions via magnetic perturbation},}\ }\href {\doibase
  10.2140/apde.2013.6.1213} {\bibfield  {journal} {\bibinfo  {journal}
  {Analysis {\&} {PDE}}\ }\textbf {\bibinfo {volume} {6}},\ \bibinfo {pages}
  {1213--1233} (\bibinfo {year} {2013})}\BibitemShut {NoStop}%
\bibitem [{\citenamefont {Fanuel}, \citenamefont {Ala\'{\i}z},\ and\
  \citenamefont {Suykens}(2017)}]{preMagCommunity2017}%
  \BibitemOpen
  \bibfield  {author} {\bibinfo {author} {\bibfnamefont {M.}~\bibnamefont
  {Fanuel}}, \bibinfo {author} {\bibfnamefont {C.~M.}\ \bibnamefont
  {Ala\'{\i}z}}, \ and\ \bibinfo {author} {\bibfnamefont {J.~A.~K.}\
  \bibnamefont {Suykens}},\ }\bibfield  {title} {\enquote {\bibinfo {title}
  {Magnetic eigenmaps for community detection in directed networks},}\ }\href
  {\doibase 10.1103/PhysRevE.95.022302} {\bibfield  {journal} {\bibinfo
  {journal} {Phys. Rev. E}\ }\textbf {\bibinfo {volume} {95}},\ \bibinfo
  {pages} {022302} (\bibinfo {year} {2017})}\BibitemShut {NoStop}%
\bibitem [{\citenamefont {Furutani}\ \emph {et~al.}(2020)\citenamefont
  {Furutani}, \citenamefont {Shibahara}, \citenamefont {Akiyama}, \citenamefont
  {Hato},\ and\ \citenamefont {Aida}}]{graphSignal2020}%
  \BibitemOpen
  \bibfield  {author} {\bibinfo {author} {\bibfnamefont {S.}~\bibnamefont
  {Furutani}}, \bibinfo {author} {\bibfnamefont {T.}~\bibnamefont {Shibahara}},
  \bibinfo {author} {\bibfnamefont {M.}~\bibnamefont {Akiyama}}, \bibinfo
  {author} {\bibfnamefont {K.}~\bibnamefont {Hato}}, \ and\ \bibinfo {author}
  {\bibfnamefont {M.}~\bibnamefont {Aida}},\ }\bibfield  {title} {\enquote
  {\bibinfo {title} {Graph signal processing for directed graphs based on the
  hermitian laplacian},}\ }in\ \href
  {https://ecmlpkdd2019.org/downloads/paper/499.pdf} {\emph {\bibinfo
  {booktitle} {Machine Learning and Knowledge Discovery in Databases}}},\
  \bibinfo {editor} {edited by\ \bibinfo {editor} {\bibfnamefont
  {U.}~\bibnamefont {Brefeld}}, \bibinfo {editor} {\bibfnamefont
  {E.}~\bibnamefont {Fromont}}, \bibinfo {editor} {\bibfnamefont
  {A.}~\bibnamefont {Hotho}}, \bibinfo {editor} {\bibfnamefont
  {A.}~\bibnamefont {Knobbe}}, \bibinfo {editor} {\bibfnamefont
  {M.}~\bibnamefont {Maathuis}}, \ and\ \bibinfo {editor} {\bibfnamefont
  {C.}~\bibnamefont {Robardet}}}\ (\bibinfo  {publisher} {Springer
  International Publishing},\ \bibinfo {address} {Cham},\ \bibinfo {year}
  {2020})\ pp.\ \bibinfo {pages} {447--463}\BibitemShut {NoStop}%
\bibitem [{\citenamefont {{Kantorovitch}}(1942)}]{Kantorovitch42}%
  \BibitemOpen
  \bibfield  {author} {\bibinfo {author} {\bibfnamefont {L.}~\bibnamefont
  {{Kantorovitch}}},\ }\bibfield  {title} {\enquote {\bibinfo {title} {{On the
  translocation of masses.}}}\ }\href {https://doi.org/10.1561/2200000073}
  {\bibfield  {journal} {\bibinfo  {journal} {{C. R. (Dokl.) Acad. Sci. URSS,
  n. Ser.}}\ }\textbf {\bibinfo {volume} {37}},\ \bibinfo {pages} {199--201}
  (\bibinfo {year} {1942})}\BibitemShut {NoStop}%
\bibitem [{\citenamefont {Bogachev}\ and\ \citenamefont
  {Kolesnikov}(2012)}]{Bogachev2012}%
  \BibitemOpen
  \bibfield  {author} {\bibinfo {author} {\bibfnamefont {V.~I.}\ \bibnamefont
  {Bogachev}}\ and\ \bibinfo {author} {\bibfnamefont {A.~V.}\ \bibnamefont
  {Kolesnikov}},\ }\bibfield  {title} {\enquote {\bibinfo {title} {The
  monge-kantorovich problem: achievements, connections, and perspectives},}\
  }\href {\doibase 10.1070/rm2012v067n05abeh004808} {\bibfield  {journal}
  {\bibinfo  {journal} {Russian Mathematical Surveys}\ }\textbf {\bibinfo
  {volume} {67}},\ \bibinfo {pages} {785--890} (\bibinfo {year}
  {2012})}\BibitemShut {NoStop}%
\bibitem [{\citenamefont {Peyr{\'{e}}}\ and\ \citenamefont
  {Cuturi}(2019)}]{Peyr2019}%
  \BibitemOpen
  \bibfield  {author} {\bibinfo {author} {\bibfnamefont {G.}~\bibnamefont
  {Peyr{\'{e}}}}\ and\ \bibinfo {author} {\bibfnamefont {M.}~\bibnamefont
  {Cuturi}},\ }\bibfield  {title} {\enquote {\bibinfo {title} {Computational
  optimal transport},}\ }\href {\doibase 10.1561/2200000073} {\bibfield
  {journal} {\bibinfo  {journal} {Foundations and Trends{\textregistered} in
  Machine Learning}\ }\textbf {\bibinfo {volume} {11}},\ \bibinfo {pages}
  {355--206} (\bibinfo {year} {2019})}\BibitemShut {NoStop}%
\bibitem [{\citenamefont {Bollob{\'a}s}\ \emph {et~al.}(2003)\citenamefont
  {Bollob{\'a}s}, \citenamefont {Borgs}, \citenamefont {Chayes},\ and\
  \citenamefont {Riordan}}]{bollobas2003directed}%
  \BibitemOpen
  \bibfield  {author} {\bibinfo {author} {\bibfnamefont {B.}~\bibnamefont
  {Bollob{\'a}s}}, \bibinfo {author} {\bibfnamefont {C.}~\bibnamefont {Borgs}},
  \bibinfo {author} {\bibfnamefont {J.}~\bibnamefont {Chayes}}, \ and\ \bibinfo
  {author} {\bibfnamefont {O.}~\bibnamefont {Riordan}},\ }\bibfield  {title}
  {\enquote {\bibinfo {title} {Directed scale-free graphs},}\ }in\ \href@noop
  {} {\emph {\bibinfo {booktitle} {Proceedings of the fourteenth annual
  ACM-SIAM symposium on Discrete algorithms}}}\ (\bibinfo {organization}
  {Society for Industrial and Applied Mathematics},\ \bibinfo {year} {2003})\
  pp.\ \bibinfo {pages} {132--139}\BibitemShut {NoStop}%
\bibitem [{\citenamefont {de~Verdi{\`{e}}re}(2013)}]{colin2013magnetic}%
  \BibitemOpen
  \bibfield  {author} {\bibinfo {author} {\bibfnamefont {Y.~C.}\ \bibnamefont
  {de~Verdi{\`{e}}re}},\ }\bibfield  {title} {\enquote {\bibinfo {title}
  {Magnetic interpretation of the nodal defect on graphs},}\ }\href {\doibase
  10.2140/apde.2013.6.1235} {\bibfield  {journal} {\bibinfo  {journal}
  {Analysis {\&} {PDE}}\ }\textbf {\bibinfo {volume} {6}},\ \bibinfo {pages}
  {1235--1242} (\bibinfo {year} {2013})}\BibitemShut {NoStop}%
\bibitem [{\citenamefont {Lieb}\ and\ \citenamefont
  {Loss}(1993)}]{lieb1993fluxes}%
  \BibitemOpen
  \bibfield  {author} {\bibinfo {author} {\bibfnamefont {E.~H.}\ \bibnamefont
  {Lieb}}\ and\ \bibinfo {author} {\bibfnamefont {M.}~\bibnamefont {Loss}},\
  }\bibfield  {title} {\enquote {\bibinfo {title} {Fluxes, laplacians, and
  kasteleyn's theorem},}\ }in\ \href {\doibase 10.1007/978-3-662-10018-9_28}
  {\emph {\bibinfo {booktitle} {Statistical Mechanics}}}\ (\bibinfo
  {publisher} {Springer Berlin Heidelberg},\ \bibinfo {year} {1993})\ pp.\
  \bibinfo {pages} {457--483}\BibitemShut {NoStop}%
\bibitem [{\citenamefont {Blum}(2012)}]{blum2012density}%
  \BibitemOpen
  \bibfield  {author} {\bibinfo {author} {\bibfnamefont {K.}~\bibnamefont
  {Blum}},\ }\href@noop {} {\emph {\bibinfo {title} {Density matrix theory and
  applications}}},\ Vol.~\bibinfo {volume} {64}\ (\bibinfo  {publisher}
  {Springer Science \& Business Media},\ \bibinfo {year} {2012})\BibitemShut
  {NoStop}%
\bibitem [{\citenamefont {Wei\ss{}e}\ \emph {et~al.}(2006)\citenamefont
  {Wei\ss{}e}, \citenamefont {Wellein}, \citenamefont {Alvermann},\ and\
  \citenamefont {Fehske}}]{KernelPolynomial2006}%
  \BibitemOpen
  \bibfield  {author} {\bibinfo {author} {\bibfnamefont {A.}~\bibnamefont
  {Wei\ss{}e}}, \bibinfo {author} {\bibfnamefont {G.}~\bibnamefont {Wellein}},
  \bibinfo {author} {\bibfnamefont {A.}~\bibnamefont {Alvermann}}, \ and\
  \bibinfo {author} {\bibfnamefont {H.}~\bibnamefont {Fehske}},\ }\bibfield
  {title} {\enquote {\bibinfo {title} {The kernel polynomial method},}\ }\href
  {\doibase 10.1103/RevModPhys.78.275} {\bibfield  {journal} {\bibinfo
  {journal} {Rev. Mod. Phys.}\ }\textbf {\bibinfo {volume} {78}},\ \bibinfo
  {pages} {275--306} (\bibinfo {year} {2006})}\BibitemShut {NoStop}%
\bibitem [{\citenamefont {Boyd}(2001)}]{Chebyshev2001}%
  \BibitemOpen
  \bibfield  {author} {\bibinfo {author} {\bibfnamefont {J.~P.}\ \bibnamefont
  {Boyd}},\ }\href
  {https://www.amazon.com/Chebyshev-Fourier-Spectral-Methods-Mathematics/dp/0486411834?SubscriptionId=AKIAIOBINVZYXZQZ2U3A&tag=chimbori05-20&linkCode=xm2&camp=2025&creative=165953&creativeASIN=0486411834}
  {\emph {\bibinfo {title} {Chebyshev and Fourier Spectral Methods: Second
  Revised Edition (Dover Books on Mathematics)}}}\ (\bibinfo  {publisher}
  {Dover Publications},\ \bibinfo {year} {2001})\BibitemShut {NoStop}%
\bibitem [{\citenamefont {Hutchinson}(1990)}]{Hutchinson1990}%
  \BibitemOpen
  \bibfield  {author} {\bibinfo {author} {\bibfnamefont {M.}~\bibnamefont
  {Hutchinson}},\ }\bibfield  {title} {\enquote {\bibinfo {title} {A stochastic
  estimator of the trace of the influence matrix for laplacian smoothing
  splines},}\ }\href {\doibase 10.1080/03610919008812866} {\bibfield  {journal}
  {\bibinfo  {journal} {Communications in Statistics - Simulation and
  Computation}\ }\textbf {\bibinfo {volume} {19}},\ \bibinfo {pages} {433--450}
  (\bibinfo {year} {1990})}\BibitemShut {NoStop}%
\bibitem [{\citenamefont {Jackson}(1912)}]{Jackson1912}%
  \BibitemOpen
  \bibfield  {author} {\bibinfo {author} {\bibfnamefont {D.}~\bibnamefont
  {Jackson}},\ }\bibfield  {title} {\enquote {\bibinfo {title} {On
  approximation by trigonometric sums and polynomials},}\ }\href {\doibase
  10.1090/s0002-9947-1912-1500930-2} {\bibfield  {journal} {\bibinfo  {journal}
  {Transactions of the American Mathematical Society}\ }\textbf {\bibinfo
  {volume} {13}},\ \bibinfo {pages} {491--491} (\bibinfo {year}
  {1912})}\BibitemShut {NoStop}%
\bibitem [{\citenamefont {Gray}(2005)}]{gray2006toeplitz}%
  \BibitemOpen
  \bibfield  {author} {\bibinfo {author} {\bibfnamefont {R.~M.}\ \bibnamefont
  {Gray}},\ }\bibfield  {title} {\enquote {\bibinfo {title} {Toeplitz and
  circulant matrices: A review},}\ }\href {\doibase 10.1561/0100000006}
  {\bibfield  {journal} {\bibinfo  {journal} {Foundations and
  Trends{\textregistered} in Communications and Information Theory}\ }\textbf
  {\bibinfo {volume} {2}},\ \bibinfo {pages} {155--239} (\bibinfo {year}
  {2005})}\BibitemShut {NoStop}%
\bibitem [{\citenamefont {Bollob\'{a}s}\ \emph {et~al.}(2003)\citenamefont
  {Bollob\'{a}s}, \citenamefont {Borgs}, \citenamefont {Chayes},\ and\
  \citenamefont {Riordan}}]{Bollobas2003}%
  \BibitemOpen
  \bibfield  {author} {\bibinfo {author} {\bibfnamefont {B.}~\bibnamefont
  {Bollob\'{a}s}}, \bibinfo {author} {\bibfnamefont {C.}~\bibnamefont {Borgs}},
  \bibinfo {author} {\bibfnamefont {J.}~\bibnamefont {Chayes}}, \ and\ \bibinfo
  {author} {\bibfnamefont {O.}~\bibnamefont {Riordan}},\ }\bibfield  {title}
  {\enquote {\bibinfo {title} {Directed scale-free graphs},}\ }in\ \href
  {http://dl-acm-org.ez67.periodicos.capes.gov.br/citation.cfm?id=644108.644133}
  {\emph {\bibinfo {booktitle} {Proceedings of the Fourteenth Annual ACM-SIAM
  Symposium on Discrete Algorithms}}},\ \bibinfo {series and number} {SODA
  '03}\ (\bibinfo  {publisher} {Society for Industrial and Applied
  Mathematics},\ \bibinfo {address} {Philadelphia, PA, USA},\ \bibinfo {year}
  {2003})\ pp.\ \bibinfo {pages} {132--139}\BibitemShut {NoStop}%
\bibitem [{\citenamefont {Shirinyan}\ \emph {et~al.}(2019)\citenamefont
  {Shirinyan}, \citenamefont {Kozin}, \citenamefont {Hellsvik}, \citenamefont
  {Pereiro}, \citenamefont {Eriksson},\ and\ \citenamefont
  {Yudin}}]{SOM2019PRB}%
  \BibitemOpen
  \bibfield  {author} {\bibinfo {author} {\bibfnamefont {A.~A.}\ \bibnamefont
  {Shirinyan}}, \bibinfo {author} {\bibfnamefont {V.~K.}\ \bibnamefont
  {Kozin}}, \bibinfo {author} {\bibfnamefont {J.}~\bibnamefont {Hellsvik}},
  \bibinfo {author} {\bibfnamefont {M.}~\bibnamefont {Pereiro}}, \bibinfo
  {author} {\bibfnamefont {O.}~\bibnamefont {Eriksson}}, \ and\ \bibinfo
  {author} {\bibfnamefont {D.}~\bibnamefont {Yudin}},\ }\bibfield  {title}
  {\enquote {\bibinfo {title} {Self-organizing maps as a method for detecting
  phase transitions and phase identification},}\ }\href {\doibase
  10.1103/PhysRevB.99.041108} {\bibfield  {journal} {\bibinfo  {journal} {Phys.
  Rev. B}\ }\textbf {\bibinfo {volume} {99}},\ \bibinfo {pages} {041108}
  (\bibinfo {year} {2019})}\BibitemShut {NoStop}%
\end{thebibliography}%

\end{document}